Noncommuting vector fields, polynomial approximations and control of inhomogeneous quantum ensemblesx

\documentclass[10pt, twocolumn,showpacs,preprintnumbers,amsmath,amssymb]{revtex4}

\usepackage{graphicx}
\usepackage{dcolumn}
\usepackage{bm}
\usepackage{epsfig}
                                                          
\begin{document}

\title{\LARGE \bf Noncommuting vector fields, polynomial approximations and control of inhomogeneous quantum ensembles}

\author{Jr-Shin Li}
\author{Navin Khaneja}
\email{navin@eecs.harvard.edu}
\homepage{http://hrl.harvard.edu/~navin}
\affiliation{Division of Applied Sciences,\\ Harvard University,
Cambridge, MA 02138}
\thanks{The work was supported by ONR 38A-1077404, AFOSR FA9550-05-1-0443 and AFOSR FA9550-04-1-0427} 

\date{\today}

\def\ep {\epsilon}


\begin{abstract}
Finding control fields (pulse sequences) that can compensate for the dispersion in the parameters governing the
evolution of a quantum system is an important problem in coherent spectroscopy and quantum 
information processing. The use of composite pulses for compensating dispersion in system dynamics is 
widely known and applied. In this paper, we make explicit the key aspects of the dynamics 
that makes such a compensation possible. We highlight the role of Lie algebras and non-commutativity in 
the design of a compensating pulse sequence. Finally we investigate three common dispersions in 
NMR spectroscopy, the Larmor dispersion, rf-inhomogeneity and strength of couplings 
between the spins.
\end{abstract}

\pacs{03.67.-a}

\maketitle

\section{INTRODUCTION}
Many applications in control of quantum systems involve controlling
a large ensemble by using the same control field. In practice, the
elements of the ensemble could show variation in the parameters 
that govern the dynamics of the system. For example, in
magnetic resonance experiments, the spins of an ensemble may have
large dispersion in their natural frequencies (Larmor dispersion), 
strength of applied rf-field (rf-inhomogeneity) and the relaxation rates of 
the spins. In solid state NMR spectroscopy of powders, the random distribution of
orientations of inter-nuclear vectors of coupled spins within an
ensemble leads to a distribution of coupling strengths \cite{Rohr}. A
canonical problem in control of quantum ensembles
is to develop external excitations that can simultaneously steer the ensemble of
systems with variation in their internal parameters from an initial
state to a desired final state \cite{Skinner1, Kobzar1, Kehlet, Skinner2}. These are called compensating pulse 
sequences as they can compensate for the dispersion in the system dynamics. From the standpoint of mathematical
control theory, the challenge is to simultaneously steer a continuum of
systems between points of interest with the same control signal. 
Typical applications are the design of excitation and inversion pulses in NMR 
spectroscopy in the presence of larmor dispersion and rf-inhomogeneity 
\cite{levitt, tyco, shaka, Garwood, Skinner1, Kobzar1, Skinner2, Pattern}
or the transfer of coherence or polarization in coupled spin ensemble with 
variations in the coupling strengths \cite{Kehlet}. In many cases of 
practical interest, one wants to find a control field that prepares the 
final state as some desired function of the parameter. For example, slice 
selective excitation and inversion pulses in magnetic resonance imaging
\cite{Silver, Rourke, Shinnar, Roux}.
The problem of designing excitations that can compensate for dispersion in the dynamics is 
a well studied subject in NMR spectroscopy and 
extensive literature exists on the subject of composite pulses 
that correct for dispersion in system dynamics \cite{levitt, tyco, shaka, Garwood}. 
The focus of this paper is not to construct a new compensating pulse sequence but rather to highlight the 
aspects of system dynamics that make such a compensation possible and give 
proofs of existence of a compensating pulse sequence. Our final goal is to understand what kind of 
dispersions can and cannot be corrected.

To fix ideas, consider an ensemble of noninteracting spin $\frac{1}{2}$ in a 
static field $B_0$ along $z$ axis and a transverse rf-field, 
$(A(t)\cos(\phi(t)), A(t) \sin(\phi(t)))$, in the $x-y$ plane. 
Let $x, y, z$ represent the coordinates of the unit vector in direction of the net magnetization vector of the ensemble. 
The dispersion in the amplitude of the rf-field is given by a dispersion parameter $\epsilon$ such that 
$A(t) = \epsilon A_0(t)$ where $\epsilon \in [1 - \delta, 1 + \delta]$, for $\delta > 0$. 
Similarly there is dispersion 
in the larmor frequency $\omega$ around a nominal value $\omega_0$, i.e., 
$\omega - \omega_0 = \Delta \omega \in [-B, B]$. In a rotating 
frame rotating with frequency $\omega_0$, the Bloch equations take the form
\begin{equation}
\label{eq:Bloch1}
\frac{d}{dt}{\left[\begin{array}{c}x\\y\\z\end{array}\right]}
=\left[\begin{array}{ccc}0&-\Delta \omega& \epsilon u(t)\\
\Delta \omega&0& -\epsilon v(t)\\-\epsilon u(t)& \epsilon v(t)&0\end{array}\right]
\left[\begin{array}{c} x\\y\\z\end{array}\right].
\end{equation}
Consider now the problem of designing controls $u(t)$ and $v(t)$
that simultaneously steer an ensemble of such systems with dispersion in their
natural frequency and strength of rf-field from an initial state
$(x,y,z)=(0,0,1)$ to a final state $(x,y,z)=(1,0,0)$
\cite{Skinner1}. This problem raises interesting questions about
controllability, i.e., showing that inspite of bounds on the strength of rf-field, 
$\sqrt{u^2(t)+v^2(t)}\leq A_{max}$, there exist excitations $(u(t), v(t))$, 
which simultaneously steer all the systems with dispersion in 
$\Delta \omega$ and $\ep$, to a ball of desired radius $r$ around the 
final state $(1,0,0)$ in a finite time (which may depend on $A_{max}$, $B$, $\delta$, and $r$). 
These are control problems 
involving infinite dimensional systems with special structure. 
Besides steering the ensemble between two points, we can ask for a control that steers an 
initial distribution of the ensemble
to a final distribution, i.e., if $X(t)$ denote the units vector $(x(t), y(t), z(t))$, consider the problem of steering an 
initial distribution $X(\Delta \omega, \ep, 0)$ to a target function $X(\Delta \omega, \ep, T)$ by 
appropriate choice of controls in equation (\ref{eq:Bloch1}). If a system with 
dispersion in parameters can be steered between states that have dependency on the 
dispersion parameter, then we say that the system is ensemble controllable with respect 
to these parameters. A more formal definition will appear later in the paper.

This paper is organized as follows. In the following section, we
introduce the key ideas and through examples, highlight the role of Lie brackets and non-commutativity in the
design of a compensating control. In section 3, we show that the Bloch equations (\ref{eq:Bloch1}), with 
bounded controls, $u(t)$ and $v(t)$ are ensemble controllable in 
the presence of Larmor dispersion and rf-inhomogeneity.  
Finally in section 4, we investigate in some generality, the notion of ensemble controllability for linear 
control systems and a class of nonlinear control systems.
 
\section{Lie Brackets and Ensemble Controllability}
\label{sec:Lie.Bracket}

\textbf{Example 1: Main Concept} {\rm To fix ideas, we begin by considering Bloch equations with only 
rf-inhomogeneity and no Larmor dispersion.  
$$ \dot{X} = \epsilon (u(t) \Omega_y + v(t) \Omega_x )X $$ where 
$$\Omega_x = \left[\begin{array}{ccc}0& 0 & 0 \\
0 & 0& -1 \\ 0 & 1 & 0 \end{array}\right],\ \ \Omega_y = \left[\begin{array}{ccc}0& 0 & 1 \\
0 & 0& 0 \\ -1 & 0 & 0 \end{array}\right], \Omega_z = \left[\begin{array}{ccc}0& -1 & 0 \\
1 & 0& 0 \\ 0 & 0 & 0 \end{array}\right] $$ are the generators of rotation around 
$x$, $y$ and $z$ axis, respectively. 

Observe for small $dt$, the evolution $ U_1(\sqrt{dt})=$ 
$$ \exp(- \epsilon \Omega_y \sqrt{dt})\ \exp(- \epsilon \Omega_x \sqrt{dt})\ \exp( \epsilon \Omega_y \sqrt{dt}) \exp( \epsilon \Omega_x \sqrt{dt})$$ to leading order in $dt$ is given by $I + (dt)[\epsilon \Omega_y, \epsilon \Omega_x]$, i.e., we can synthesize the generator $[\epsilon \Omega_x, \epsilon \Omega_y] = \epsilon^2 \Omega_z,$ by back and 
forth maneuver in the directly accessible directions $\Omega_x$ and $\Omega_y$. 

Similarly, the leading order term in the evolution 
$$ U_2 = U_1(-\sqrt{dt})\exp(-\epsilon \Omega_y dt) U_1(\sqrt{dt})\exp(\epsilon \Omega_y dt).$$ is
$[\ep \Omega_y, [\ep \Omega_x, \ep \Omega y]] = {\ep}^3 \Omega_x.$ Therefore by successive Lie brackets, we can 
synthesize terms of the type ${\ep}^{2k+1} \Omega_x$. Now using $\{\ep \Omega_x, \epsilon^3 \Omega_x, \dots,
\epsilon^{2n+1} \Omega_x \}$ as generators, we can produce an evolution 
$$\exp \{ \sum_{k=0}^n c_k \ep^{2k+1} \Omega_x \}, $$ where $n$ and the coefficient $c_k$ can be chosen so 
that $$\sum_{k=0}^n c_k \ep^{2k+1} \approx \theta $$ for all $\ep \in [1-\delta, 1 + \delta]$.  Hence we can 
generate an evolution $\exp(\theta \Omega_x)$ for all $\ep$ to any desired accuracy.
Therefore, we achieve robustness with dispersion to $\ep$ by generating suitable Lie brackets. 
Similar arguments show that we can generate any evolution $\exp(\beta \Omega_y)$ and as a result 
any three dimensional rotation in a robust way. It is also now easy to see that we can synthesize 
rotation $\Theta$ with a desired functional dependency on the parameter $\epsilon$. Parametrize a 
rotation in $\Theta \in SO(3)$ by the Euler angles $(\alpha, \beta, \gamma)$ such that
$\Theta = \exp(\alpha \Omega_x) \exp(\beta \Omega_y) \exp(\gamma \Omega_x)$. Given continuous function
 $(\alpha(\ep), \beta(\ep), \gamma(\ep))$, of $\ep$, we can find polynomials 
that approximate $\alpha(\epsilon)$, $\beta(\ep)$ and $\gamma(\ep)$ arbitrarily well and use these to generate
a desired rotation $\Theta(\ep)$ as a function of $\ep$. Hence there exists a control field that maps a smooth 
initial distribution $X_{\ep}(0)$ to a target distribution $X_{\ep}^F$. 

\textbf{Remark:} Note we have assumed that $\ep >0$. 
The above system will fail to be ensemble controllable if 
$\ep \in [-\ep_0, \ep_0]$, as we cannot approximate an even function $f(\ep)= \theta$, with an odd 
degree polynomial.

\textbf{Remark} The key idea in designing compensating pulse sequence is to 
synthesize higher order Lie brackets that raise the dispersion parameters to higher 
powers. The various powers of the dispersion parameter can be combined for compensation as 
explained above. The construction presented here is not the most efficient 
way of achieving a desired level of compensation. The construction given here however 
presents in a transparent way the role of higher order lie bracketing. 
We now consider an example when there are more than one parameter in the system dynamics.

\textbf{Example 2} Now consider the system
$$ \dot{X} = (\epsilon_1 u(t) \Omega_x + \epsilon_2 v(t) \Omega_y )X. $$ 
where $\ep_1 \in [1 - \delta_1, 1 + \delta_1]$ and $\ep_2 \in [1 - \delta_2, 1 + \delta_2]$, 
for $0 < \delta_1 < 1$ and  $0 < \delta_2 < 1$. The system is ensemble controllable with 
respect to dispersions $\epsilon_1$ and $\epsilon_2$.

The reasoning proceeds along the same lines as before except now we have two dispersions 
parameters that are independent. 
Let $ad_X(Y)$ represent the lie bracket $[X,Y]$ (similarly $ad_X^2(Y)=[X,[X,Y]]$). Consider 
the identity,
\begin{equation}
ad_{\ep_1 \Omega_x}^{2k + 1} (\ep_2 \Omega_y) = (-1)^{k} {\ep_1}^{2k+ 1}\ep_2 \Omega_z. \\
\end{equation} for $k=0,1,2, \dots, n$.

We can now choose coefficients $c_k$ such that $\sum_k c_k \ep_1^{2k + 1}$ approximates a constant 
function
over the range of $\ep_1$. As a result, we can generate the bracket direction 
$\ep_2 \Omega_z$. Now using the bracket directions $\ep_2 \Omega_z$ and $\ep_2 \Omega_y$, and the construction in 
Example 1, we can further compensate the dispersion of $\ep_2$ and steer the whole ensemble together to a 
desired point. Infact the 
final point can be made to depend explicitly on $\epsilon_1$ and $\epsilon_2$ by synthesizing the bracket 
directions $\sum_{kl} (c_{kl} \ep_1^{2k + 1}\ep_2^{2l+1}) \Omega_z $, and 
$\sum_{kl} (d_{kl} \ep_1^{2k}\ep_2^{2l+1}) \Omega_y$. The coefficients
$c_{kl}$ and $d_{kl}$ can be now so chosen that we can approximate
rotations $\exp(\theta(\ep_1, \ep_2)\Omega_x)$ and  $\exp(\theta(\ep_1, \ep_2)\Omega_y)$. Therefore we 
have ensemble controllability.

\textbf{Example 3: Phase dispersions cannot be compensated} Consider an ensemble of Bloch equations
\begin{equation}\label{eq:phase}
\dot{X_{\theta}} = A(t)(\cos(\phi(t) + \theta) \Omega_x + \sin(\phi(t) + \theta) \Omega_y )X_{\theta},
\end{equation} 
where there is dispersion in the phase of the rf field. The system is not ensemble controllable with respect 
to the dispersion $\theta \in [\theta_1, \theta_2]$.

{\it Proof:} The simplest way to see this is to make the change of co-ordinates 
$Y_{\theta} = \exp(- \Omega_z \theta) X_{\theta}$. The resulting system then takes the form
$$ \dot{Y_{\theta}} = A(t)(\cos(\phi(t)) \Omega_x + \sin(\phi(t)) \Omega_y )Y_{\theta}. $$
Since all $Y_{\theta}$ see the same field, they have identical trajectories. As a 
result $X_{\theta}$ cannot be simultaneously steered from $(0, 0, 1)$ to $(1, 0, 0)$.
Lack of ensemble controllability can also be understood by looking at Lie brackets of the generators. Equation
(\ref{eq:phase}) can be written as 
$$\dot{X_{\theta}}=\{A(t)\cos(\phi(t))B_1 + A(t)\sin(\phi(t)) \}X_{\theta}, $$where
the $B_1 = \cos(\theta) \Omega_x + \sin(\theta) \Omega_y$
and $B_2 = -\sin(\theta) \Omega_x + \cos(\theta) \Omega_y$. 
Observe that $B_3 = [B_1, B_2] = \Omega_z$. Therefore, all iterated brackets of $B_i's$ are linear 
in $\cos(\theta)$ and $\sin(\theta)$ and we cannot raise the 
dispersion parameters  $\cos(\theta)$ and $\sin(\theta)$ to higher 
powers and therefore cannot compensate for the dispersion in $\theta$. 

\textbf{Example 4: Larmor dispersion in the presence of strong rf-field} Now consider the Bloch equations
$$ \dot{X_{\theta}} = (\omega \Omega_z + u(t) \Omega_x + v(t) \Omega_y )X_{\theta}.$$ 
with dispersion in the Larmor frequencies. The system is ensemble controllable with respect to the 
dispersion parameter $\omega$.

Note because of the assumption of strong fields, we can reverse the evolution of the drift term
\begin{equation}\label{eq:pirot}
\exp(\pi \Omega_x)\exp(\omega \Omega_z dt)\exp(-\pi \Omega_x) = \exp(-\omega \Omega_z dt).
\end{equation}
Now as before a maneuver
$$ \exp(-\omega \Omega_z \sqrt{dt})\ \exp(-\Omega_x \sqrt{dt})\ \exp( \omega \Omega_z \sqrt{dt}) 
\exp( \Omega_x \sqrt{dt})$$ produces the bracket direction 
$[\omega \Omega_z, \Omega_x]= \omega \Omega_y$ 
to leading order. Similarly 
$ [\omega \Omega_z [\omega \Omega_z, \Omega_x]] = -(\omega)^2 \Omega_x $.
Hence, we can generate higher brackets with even and odd powers of $\omega$. To see that 
the system is ensemble 
controllable consider the Lie bracket relation 
$ad_{(\omega \Omega_z)^{2n}} \Omega_x = (-1)^n \omega^{2n} \Omega_x$ and   
$ad_{(\omega \Omega_z)^{2n+1}} \Omega_y = (-1)^{n+1} \omega^{2n+1} \Omega_x$, we can synthesize 
an evolution $\exp(\sum_k c_k \omega^k \Omega_x)$ and  similarly the evolution 
$\exp(\sum_k d_k \omega^k \Omega_y)$. The coefficients $c_k$ and $d_k$ can be chosen to 
approximate Euler angles $(\alpha(\omega), \beta(\omega), \gamma(\omega))$ and we therefore as in Theorem 1, 
have ensemble controllability.

\textbf{Remark}  Note if we have only one quadrature of the control field i.e.,  
$$ \dot{X} = (\omega \Omega_z + u(t) \Omega_x)X, $$ then we can only synthesize 
the generator $\Omega_y$ with 
odd powers of $\omega$, with $\omega \in [-B, B]$. Therefore an evolution of the 
form $\exp(f(\omega) \Omega_y)$ cannot be 
approximated if $f$ is an even function. 

\textbf{Example 5: Dispersion in Coupling Strengths} Consider two coupled qubits with Ising type 
interactions with dispersion in coupling strengths J. The interaction 
Hamiltonian $H_c = J \sigma_{1z}\sigma_{2z}$, with $J \in J_0 [1-\delta, 1 + \delta]$, $\delta > 0$. 
Although not necessary, for simplicity of exposition, we assume
that we can produce local unitary transformation on the qubits much faster than the evolution of couplings. We now 
show that it is possible to compensate for dispersion in $J$ and generate any quantum logic with high fidelity. 

By local transformations we can synthesize the effective Hamiltonian 
$$ J \sigma_{1y}\sigma_{2z} = \exp(i\sigma_{1x}\frac{\pi}{2})(J \sigma_{1z}\sigma_{2z})\exp(-i\sigma_{1x}\frac{\pi}{2}).$$ 
Now using $B_1 = -i 2 \sigma_{1y}\sigma_{2z}$ and $B_2 = -i 2 \sigma_{1z}\sigma_{2z}$ as generators 
we get $[J B_1 [J B_1, J B_2] = -J^3 B_2$. Now using a construction similar to one in example 1, we can synthesize the evolution 
$\exp(\sum_kc_kJ^{2k+1}\sigma_{1z}\sigma_{2z})$, where the coefficients $c_k$ are chosen such that 
$\sum_kc_kJ^{2k+1}\approx J_0$ over the range of dispersion of $J$. Hence we have compensate for dispersion in $J$.
We also have ensemble controllability with respect to the parameter $J$. Let
$$A(J) = \exp(-i (a(J) \sigma_{1x}\sigma_{2x} + b(J) \sigma_{1y}\sigma_{2y} + c(J) \sigma_{1z}\sigma_{2z})). $$ We can write an arbitrary two qubit gate with the dependency on $J$ as 
$$U_2(J)\otimes U_1(J) \ A(J) \ V_2(J) \otimes V_1(J).$$ where $U_1, V_1$ and $U_2, V_2$ are local unitaries on qubits 
$1$ and $2$ respectively. We can 
synthesize them with a explicit dependence on $J$ as follows. Using the commutation relations  
of the type $[-iJ 2 \sigma_{1y} \sigma_{1z}, -i J 2 \sigma_{1z} \sigma_{1z}]= -iJ^2 \sigma_{1x}$, we can synthesize
generators $-i(J^2)^k \sigma_{1x}$, $-i(J^2)^k \sigma_{1y}$, $-i(J^2)^k \sigma_{2x}$, 
$-i(J^2)^k \sigma_{2y}$ ($k=0, 1, 2, \dots$ ) and use these to synthesize $U_1(J), V_1(J), U_2(J), V_2(J)$. 

\textbf{Remark} Using similar ideas as above, it is possible to compensate for more general coupling tensor. 
Consider the coupling tensor 
$$ \alpha \sigma_{1x}\sigma_{2x} + \beta \sigma_{1y}\sigma_{2y} + \gamma \sigma_{1z}\sigma_{2z}. $$ 
with dispersion in $\alpha, \beta, \gamma$. Now observe for $U= \exp(-i \pi \sigma_x)$, and
$A =\exp(-i (\alpha \sigma_{1x}\sigma_{2x} + \beta \sigma_{1y}\sigma_{2y} + \gamma \sigma_{1z}\sigma_{2z}))$,  
$$U A U^{\dagger} A =  \exp(-i \gamma 2 \sigma_{1z}\sigma_{2z}). $$ So we only need to take care of the dispersion in $\gamma$ and the construction is similar to the one before.

\section{Ensemble controllability of the Bloch Equations with bounded controls}

We consider again the system (\ref{eq:Bloch1}) but now with bounded controls, so that we 
cannot produce rotations of the type $\exp(-\Omega_x \pi)$ in arbitrarily small time as in equation  
(\ref{eq:pirot}). 
Nonetheless the system is still ensemble controllable as shown below. 
Our construction initially follows the well known algorithm of Shinnar-Roux \cite{Shinnar, Roux}. We then show how this construction can be extended 
to show ensemble controllability with respect to larmor dispersion and rf-inhomogeneity in Bloch equations. The solution to the Bloch equation (\ref{eq:Bloch1}) is a rotation
$$X(T)=RX(0),$$ where $R\in SO(3)$. We work with $SU(2)$ representation of these rotations. Recall a rotation by angle $\phi$ around the unit vector $(n_x, n_y, n_z)$ has a $SU(2)$ representation of the form

\begin{eqnarray}
U=\left[\begin{array}{cc}\alpha&-\beta^{*}\\
\beta&\alpha^{*}\end{array}\right],
\end{eqnarray}
where $\alpha$ and $\beta$ are the Cayley-Klein parameters satisfying
\begin{eqnarray}
\label{eq:CK1}
\alpha&=&\cos\frac{\phi}{2}-in_{z}\sin\frac{\phi}{2},\\
\label{eq:CK2} \beta&=&-i(n_x+in_y)\sin\frac{\phi}{2}, \\
\label{eq:determinant}&&\alpha\alpha^{*} + \beta\beta^{*}=1.
\end{eqnarray}The Bloch equation then takes the form
$$\dot{U} = -\frac{i}{2} \left[\begin{array}{cc}\omega & u-iv \\
u + iv & -\omega \end{array}\right] U. $$ 

The  rotation $U$ is simply represented by its first column (also termed 
spinor representation) $\psi=\left[\begin{array}{c}\alpha\\ \beta\end{array}\right]$.
We first consider piecewise-constant controls $u(t)$ and $v(t)$. The
net rotation under these controls can be represented as successive
rotations
\begin{eqnarray}
\label{eq:SU2rots} U=U_{n}U_{n-1}\ldots U_{1}U_{0},\nonumber
\end{eqnarray}
where $U_j=\left[\begin{array}{cc}a_j&-b_{j}^{*}\\
b_j&a_{j}^{*}\end{array}\right]$ and $a_j$, $b_j$ are the
Cayley-Klein parameters for the $j$th interval. Defining the
multiplication of the matrices $U_j$ up to $k$ by
$$\left[\begin{array}{cc}\alpha_k&-\beta_k^{*}\\
\beta_k&\alpha_k^{*}\end{array}\right]=\left[\begin{array}{cc}a_k&-b_{k}^{*}\\
b_k&a_{k}^{*}\end{array}\right]\ldots\left[\begin{array}{cc}a_{0}&-b_{0}^{*}\\
b_{0}&a_{0}^{*}\end{array}\right],$$ the effect of the controls can
then be calculated by propagating the spinor
\begin{eqnarray}
\label{eq:spinor} \left[\begin{array}{c}\alpha_{k}\\
\beta_{k}\end{array}\right]=\left[\begin{array}{cc}a_{k}&-b_{k}^{*}
\\b_{k}&a_{k}^{*}\end{array}\right]\left[\begin{array}{c}\alpha_{k-1}\\
\beta_{k-1}\end{array}\right]
\end{eqnarray}
with the initial condition $\left[\begin{array}{c}\alpha_{0}\\
\beta_{0}\end{array}\right]
=\left[\begin{array}{c}1\\0\end{array}\right]$. The duration $\Delta t$, 
over which the controls $u$ and $v$ are constant can be chosen small enough such that, the net rotation
can be decomposed into two sequential rotations since
$$e^{(\omega\Omega_{z}+u\Omega_y-v\Omega_x)\Delta
t}\approx e^{(u\Omega_y-v\Omega_x)\Delta
t}\,e^{\omega\Omega_{z}\Delta t}.$$ 
Under this assumption, we can write the rotation $U_k$ as a rotation
around $z$-axis by an angle $\omega\Delta t$ followed by a rotation
about the applied control fields by an angle $\phi_k$ in $SU(2)$
representation
\begin{eqnarray}
\label{eq:hardpulse} U_{k}=\left[\begin{array}{cc}C_k&-S_{k}^{*}
\\S_{k}&C_{k}\end{array}\right]\left[\begin{array}{cc}z^{1/2}&0
\\0&z^{-1/2}\end{array}\right],
\end{eqnarray}
where
\begin{eqnarray}
 C_k&=&\cos\frac{\phi_{k}}{2},\qquad\qquad
S_k\,\,=\,\,-ie^{i\theta_{k}}\sin\frac{\phi_{k}}{2},\\
\label{eq:CkSk} \phi_k&=&A_k \Delta t,\qquad\quad\quad\theta_k\,\,=\,\,\tan^{-1}\frac{v_k}{u_k},
\nonumber\\
A_k&=&\sqrt{u_k^2+v_k^2},\qquad\qquad\quad
z\,\,=\,\,e^{-i\omega\Delta t}.\nonumber
\end{eqnarray}
Plugging (\ref{eq:hardpulse}) into (\ref{eq:spinor}), we get the
recursion relation of the spinor
\begin{eqnarray}
\label{eq:recursion1} \left[\begin{array}{c}\alpha_{k}\\
\beta_{k}\end{array}\right]
=z^{\frac{1}{2}}\left[\begin{array}{cc}C_k&-S_{k}^{*}z^{-1}
\\S_{k}&C_{k}z^{-1}\end{array}\right]\left[\begin{array}{c}\alpha_{k-1}\\
\beta_{k-1}\end{array}\right].\nonumber
\end{eqnarray}
Defining $P_k=z^{-k/2}\alpha_k$ and $Q_k=z^{-k/2}\beta_k$, the
recursion may then be reduced to
\begin{eqnarray}
\label{eq:recursion1} \left[\begin{array}{c}P_{k}\\
Q_{k}\end{array}\right]=\left[\begin{array}{cc}C_k&-S_{k}^{*}z^{-1}
\\S_{k}&C_{k}z^{-1}\end{array}\right]\left[\begin{array}{c}P_{k-1}\\
Q_{k-1}\end{array}\right]
\end{eqnarray}
with the initial condition
\begin{eqnarray}
\label{eq:initialspinor} \left[\begin{array}{c}P_{0}\\
Q_{0}\end{array}\right]
=\left[\begin{array}{c}1\\0\end{array}\right].
\end{eqnarray}
Having the recursion (\ref{eq:recursion1}) and the initial condition
(\ref{eq:initialspinor}), the spinor at the $n$th time step can be
represented as the $(n-1)$-order polynomials in $z$ (the parameter $z$ encodes the 
dispersion parameter $\omega$. 
\begin{eqnarray}
\label{eq:poly1} P_n(z)&=&\sum_{k=0}^{n-1}\,p_{k}z^{-k},\\
\label{eq:poly2} Q_n(z)&=&\sum_{k=0}^{n-1}\,q_{k}z^{-k}.
\end{eqnarray}
Note that
\begin{eqnarray}
\label{eq:polynorm} |P_n(z)|^2+|Q_n(z)|^2=1,
\end{eqnarray}
which follows from (\ref{eq:determinant}). The representation of 
rotation produced by the controls $u(t)$ and $v(t)$ has now been
reduced from a product of $n$ matrices in $SU(2)$ to two
$(n-1)$-order polynomials. The desired final states of an ensemble
of systems in (\ref{eq:Bloch1}), described by Cayley-Klein
parameters, are two functions of $z$, and hence of $\omega$.
We can now design two polynomials $P_n(z)$
and $Q_n(z)$ such that we can approximate any 
desired smooth functions $F_{\alpha}(z)$ and $F_{\beta}(z)$
satisfying $ |F_{\alpha(z)}|^2+|F_{\beta}(z)|^2=1$, which characterizes 
the desired spinor we want as function of $z$. Now we can work backwards and 
compute the $u_k's$ and $v_k's$ that will produce $P_n(z)$ and $Q_n(z)$. 
Note by multiplying both sides of
(\ref{eq:recursion1}) by the inverse of the rotation matrix we get 
\begin{eqnarray}
\label{eq:backrecursion} \left[\begin{array}{c}P_{k-1}\\
Q_{k-1}\end{array}\right]=\left[\begin{array}{c}C_kP_k+S_{k}^{*}Q_k
\\(-S_{k}P_{k}+C_kQ_k)z\end{array}\right],
\end{eqnarray}
and the constraint of (\ref{eq:polynorm}) is still preserved. We have a 
backward recursion where we use the knowledge of coefficients of $P_k(z)$ and 
$Q_k(z)$ to compute $P_{k-1}(z)$ and $Q_{k-1}(z)$. This is the well known 
Shinnar Roux \cite{Shinnar, Roux} algorithm. Because $P_{k-1}(z)$ and $Q_{k-1}(z)$ are
lower order polynomials, the leading term in $P_{k-1}$ and the
low-order term in $Q_{k-1}(z)$ must drop out
\begin{eqnarray}
\label{eq:highorder} C_{k}P_{k,k-1}+S_{k}^{*}Q_{k,k-1}&=&0,\\
\label{eq:lowerorder} -S_kP_{k,0}+C_{k}Q_{k,0}&=&0,
\end{eqnarray}
where $P_{k,m}$ denotes the coefficient of $z^{-m}$ term in
$P_k(z)$. Observe that these two equations are equivalent as may be
seen by expanding (\ref{eq:polynorm}) as a polynomial,
$${P}(z)=\sum_{m=0}^{n-1}\sum_{i=0}^{m}\left[P_{i}P_{m-i}^{*}
+Q_{i}Q_{m-i}^{*}\right]z^{-m}=1,$$ and noting that all but the
constant term are zero. The coefficient of $z^{-(k-1)}$ in
${P}(z)$ gives
$$P_{k,k-1}P_{k,0}^{*}+Q_{k,k-1}Q_{k,0}^{*}=0.$$
With this relation either equation (\ref{eq:highorder}) or
(\ref{eq:lowerorder}) may be derived from the other. Choosing
(\ref{eq:lowerorder}) and combining it with (\ref{eq:CkSk}), we get
\begin{eqnarray}
\label{eq:QPratio}
\frac{Q_{k,0}}{P_{k,0}}=\frac{-ie^{i\theta_k}\sin\frac{\phi_k}{2}}
{\cos\frac{\phi_k}{2}}.
\end{eqnarray}
This gives the rotation angle
\begin{eqnarray}
\label{eq:rotangle}\phi_k=2\tan^{-1}\Big|\frac{Q_{k,0}}{P_{k,0}}\Big|.
\end{eqnarray}
Combining (\ref{eq:QPratio}) and (\ref{eq:rotangle}), we obtain the
phase of the controls
$$\theta_k=\measuredangle\Big(\frac{iQ_{k,0}}{P_{k,0}}\Big).$$
The controls $u_k$ and $v_k$ are then
\begin{eqnarray}
\label{eq:flip} u_k&=&\frac{\phi_k}{\Delta t}\sin\theta_k,\\
 v_k&=&\frac{\phi_k}{\Delta t}\cos\theta_k,\nonumber.
\end{eqnarray}
These expressions for controls coupled with the inverse recursion in
(\ref{eq:backrecursion}) construct the piecewise constant controls
$u_k, v_k$ that generate polynomial approximations $P_n(z)$ and $Q_n(z)$ of the target
function $F_{\alpha}(z)$ and $F_{\beta}(z)$. 

In particular, if we choose $Q_n(z) = -i\sin\frac{\phi}{2}$ and
$P_n(z) = \cos\frac{\phi}{2}$, we obtain a broadband rotation around $x$ axis 
by angle $\phi$ and similarly by choosing $Q_n(z) = \sin\frac{\phi}{2}$ and
$P_n(z) = \cos\frac{\phi}{2}$, we obtain an approximation to a 
broadband rotation around $y$ axis by angle $\phi$. 

If the amplitude of the controls is bounded, we can choose $\phi$ small 
enough so that it can be achieved by small flip angles $\phi_k$ in equation
(\ref{eq:flip}). Now we can concatenate these rotations to achieve a rotation 
with a bigger angle and thereby maintain the bounds on the control.

Now we consider the case when there is also rf-inhomogeneity. If we produce a small flip angles $\phi$ compensating for dispersion in $\omega$, then dispersion 
$\ep$ in the strength of the control $u$ and $v$, results in  
$Q_n(z, \ep) \approx -ie^{i\theta_k}\frac{\phi}{2}\ep$ and the ensemble executes an
effective rotation $\exp(-i\ep \frac{\phi}{2} (\cos(\theta_k) \sigma_x + \sin(\theta_k) \sigma_y ))$.
Now using methods of example one, we can concatenate many such rotations to compensate for $\ep$.
We now show ensemble controllability with respect to both dispersion in the natural 
frequency $\omega$ and strength of the rf-field. 

We again write the final rotation $U \in SU(2)$ as
$$U = U_nU_{n-1}\dots U_1U_0, $$ where 
\begin{eqnarray*}
U_{k}(z,\ep) &=&\left[\begin{array}{cc}C_k(\ep)&-S_{k}(\ep)^{*}
\\S_{k}(\ep)&C_{k}(\ep)\end{array}\right]\left[\begin{array}{cc}z^{1/2}&0
\\0&z^{-1/2}\end{array}\right].
\end{eqnarray*}Note that the flip angle has a dependence on the parameter $\ep$.
We can now choose a desired $Q_n(z,\ep)= \sum_{k=0}^{n-1}q_k(\ep)z^{-k}$ and $P_n(z,\ep)= \sum_{k=0}^{n-1}p_k(\ep)z^{-k}$ and find the flip angles
$(\theta_k(\ep), \phi_k(\ep))$ that creates these polynomials. Now we can use the 
results of Example 1 to find pulse sequences that will synthesize 
$(\theta(\ep), \phi(\ep))$. This then establishes the ensemble controllability with 
respect to both $\omega$ and $\ep$. Such constructions can also be used to generate pattern pulses
that selectively excite the Bloch equations 
with parameters lying in a given subset of $\omega-\ep$ space \cite{Pattern}. 

\hfill $\Box$ \\

We now investigate the subject of ensemble control from a general control theory perspective. 

\section{Ensemble Controllability}

Consider a family of control systems
\begin{equation}
\label{eq:general}
\frac{dx_s}{dt}= f_s(x_s, u, t),
\end{equation}
indexed by the parameter vector $s$ taking values in some compact set
$\Omega\subset\mathbf{R^{d}}$. The same
control $u(t) \in \mathbf{R^{m}}$ is being 
used to simultaneously steer this family of control systems. For such
systems, we define the notion of ensemble controllability as following. \\ \\

\textbf{Definition 1} The family of systems in (\ref{eq:general}) is called
\emph{ensemble controllable}, if there exists a control law
${u(t)}$ such that starting from any initial state $x_s(0)$, the system 
can be steered to 
within a ball of radius $\epsilon$ around
the target state $g_s$, i.e. $\parallel x_s(T)-g_s \parallel <\epsilon$. 
Here $x_s(0)$ and $x_s(T)$ are
interpreted as functions of the variable $s$ at time 0 and $T$ and
$\parallel\cdot\parallel$ denotes a desired norm on function space. In this paper we 
take $\parallel\cdot\parallel$ to denote $L_{2}$ norm. 
The final time $T$ may depend on $\epsilon$. \\ \\

\textbf{Definition 2} An ensemble of systems is called \emph{point
ensemble controllable} if both the initial state $x_s(0)$ and the 
target state $g_s$ are constant functions and the ensemble of systems
can be steered between the initial and final states as defined above. \\ 

The key problem of interest in ensemble controllability is to characterize
the properties of the ensemble such that if each system of the ensemble is controllable, the 
whole family is ensemble controllable. We begin with considering linear systems
$\dot{x} = Ax + Bu$, where $B$ is a $n\times m$ matrix and $u \in {R^{m}}$

\textbf{Theorem 1}({\bf A negative result:}) An ensemble of linear control systems 
is not ensemble controllable if there is a variation in the control matrix $B$.

{\it Proof:} Consider the family of systems 
$$ \frac{dx_s}{dt}= (A x_s + B_s u). $$ By  variation of constant formula
$$ x_s(T) = \exp(AT)x_s(0) + \exp(AT)\int_0^{T}\exp(-A\tau)B_su(\tau)d \tau. $$ Since
$x_s(T)$ is linear in control, variation in $B$ in general makes the system 
ensemble uncontrollable. 

\textbf{Remark} Observe $B_su = \sum_{k=1}^m u_k b_s^k$ where $b^k$ are the columns of 
$B$. we can think of $b_k$ as constant vector fields that generate translations. Since
$b_k$ all commute, their Lie brackets do not generate terms carrying higher powers
of the dispersion parameters.

\textbf{Theorem} Consider an ensemble $(A,b)_s$, of linear single input controllable systems 
\begin{equation}\label{eq:linear1}
\dot{x} = A_sx  + ub_s.
\end{equation} The system is ensemble controllable only if distinct $A_s$ have 
distinct eigenvalues and $A_s$ is full rank over the ensemble.

{\it Proof:} Since each of the systems is controllable, there exists a similarity transformation 
$T_s$ such that

\begin{equation}\label{eq:conform}
A_s = T_s^{-1}\left[\begin{array}{ccccc}0 & 1 & 0 & \dots & 0 \\
0 & 0 & 1 & 0 & 0 \\ \vdots & \vdots & \ddots & \ddots & \vdots \\
0 & 0 & \dots & 0 & 1 \\
-a_0^s & -a_1^s & \dots & -a_{n-2}^s & a_{n-1}^s \end{array}\right]T_s 
\end{equation} and 
\begin{equation}
b = T_s^{-1}\left[\begin{array}{c} 0 \\ 0 \\ 0 \\ 0 \\ 1 \end{array} \right] 
\end{equation}
where $a_i^s$ are the coefficients of the characteristic polynomial of $A_s$ and 
show dispersion over the 
ensemble. First observation is that for the ensemble to be controllable, we must have $T_s$ to be a 
explicit function of $(a_0^s, a_1^s, a_2^s, \dots, a_{n-1}^s)$. Suppose two distinct 
$T_1$ and $T_2$ correspond 
to the same $(a_0, a_1, a_2, \dots, a_{n-1})$, then starting from $x=0$, for equation (\ref{eq:conform}), 
the final points $x_1$ and $x_2$ 
for the two systems at any  time $t$ are related by $x_1(t) = T_1T_2^{-1}x_2(t)$. Hence there 
exists no 
control that steers the two systems to the same point starting from  $0$. Therefore $T_s$
should be a function of $(a_0^s, a_1^s, a_2^s, \dots, a_{n-1}^s)$, i.e. no two distinct $A$ have the same 
characteristic polynomial. We now show the necessity of $A$ to be full rank. Observe if $A$ 
is not full rank, 
i.e., $a_0^s = 0$, then $A^kb$ for $k>0$ always 
lies in the subspace $H = span \{ A^ib \}_{i=1}^{n-1}$. Therefore the ensemble starting 
from $x=0$, cannot be driven to a final state $f(a_1^s, a_2^s, \dots, a_{n-1}^s)b$, 
where $f$ is an arbitrary smooth scalar function of of $(a_1^s, a_2^s, \dots, a_{n-1}^s)$. 

\textbf{Remark} As mentioned before the key idea in designing compensating pulse sequence is to 
synthesize higher order Lie brackets that raise the dispersion parameters to higher 
powers. As a results nilpotent control systems are not ensemble controllable as we cannot generate a 
desired higher power of the dispersion parameter. Consider the control system
$$ \dot{x} = \sum_i u_i(t) \epsilon_i g_i(x). $$  
If the Lie algebra generated by $g_i$ is nilpotent, the system is not ensemble controllable.

\textbf{Example} Consider the well studied nonholonomic integrator
$$ \frac{d}{dt} \left[\begin{array}{c}x_1 \\ x_2 \\ x_3 \end{array}\right] = u_1 \epsilon \left[\begin{array}{c} 1 \\ 0 \\ - x_2  \end{array}\right] + u_2 \epsilon \left[\begin{array}{c} 0 \\ 1 \\ x_1  \end{array}\right].$$ The system is 
not ensemble controllable with respect to the parameter $\epsilon$. The control vector fields $g_1 = \frac{\partial}{\partial x_1} - x_2 \frac{\partial}{\partial x_3}$ and 
$g_2 = \frac{\partial}{\partial x_2} - x_1 \frac{\partial}{\partial x_3}$ generate a nilpotent algebra 
the Heisenberg algebra. Observe that
$[\epsilon g_1, \epsilon g_2] = 2 \epsilon^2 \frac{\partial}{\partial x_3}$, which commutes with everything else.

\textbf{Remark} {\rm Let $G$ be a semi-simple Lie group and let $X \in G$. Consider the control system 
$$\dot{X} = (\sum_i u_i \ep_i B_i)X, $$ such that $\ep_i$ have dispersion in the range $[1-\delta_i, 1+ \delta_i]$,for $\delta_i >0$. If Lie algebra generated by $\{B_i\}$ spans the tangent space of $G$, then the system is ensemble controllable.
The construction is similar in spirit to example 2.

In this paper, we have tried to motivate the study of problems involving control of ensembles of dynamical systems with dispersion in the parameters. Such problems arise naturally in areas of coherent spectroscopy, quantum information processing and control of quantum systems in general. We have tried to make explicit the role of noncommutativity as a key aspect of the dynamics that makes design of a compensating control signal possible. We note again that the constructions given in this paper donot provide the most efficient schemes for compensation, yet they illustrate the main ideas of the paper in a transparent way. The subject of ensemble control appears to be very rich and we anticipate that many interesting results of both theoretical and practical importance will arise from a systematic study in this field.



\end{document}